\documentclass[conference]{IEEEtran}
\IEEEoverridecommandlockouts

\usepackage[utf8]{inputenc}
\usepackage{cite}
\usepackage{overpic}
\usepackage{amsmath,amssymb,amsfonts}
\usepackage{bm}
\usepackage{graphicx}
\usepackage{textcomp}
\usepackage{xcolor}
\usepackage{float}
\usepackage{amsthm}
\usepackage{graphicx}
\usepackage{epstopdf}
\usepackage{amsmath,bm,bbm}
\usepackage{amsfonts}
\usepackage{amssymb}
\usepackage{color}
\usepackage{multirow}
\usepackage{multicol}
\usepackage{soul,xcolor}
\usepackage{setspace}
\usepackage{algorithm}
\usepackage{algpseudocode}
\usepackage[normalem]{ulem}

\definecolor{mypurple}{HTML}{9933FF}
\definecolor{mygreen}{HTML}{009900}

\definecolor{orange}{RGB}{0,112,192}
\theoremstyle{plain}

\newtheorem{remark}{Remark}

\newcommand{\mathacr}[1]{\mathsf{#1}}

\newcommand{\vect}[1]{\mathbf{#1}}

\def\Ttran{\mbox{\tiny $\mathrm{T}$}}
\begin{document}
\makeatletter
\newcommand*{\rom}[1]{\expandafter\@slowromancap\romannumeral #1@}
\makeatother

\title{ \huge Rethinking Energy Efficiency in Cell-Free Massive MIMO: The Role of Processing and Optical Fronthaul 
}
\author{\IEEEauthorblockN{Ozan Alp Topal$^{\dagger}$, \"Ozlem Tu\u{g}fe Demir$^*$,  and Cicek Cavdar$^{\dagger}$}\\
	\IEEEauthorblockA{ {$^\dagger$Department of Communication Systems, KTH Royal Institute of Technology, Stockholm, Sweden
		}  \\ {$^*$Department of Electrical and Electronics Engineering, Bilkent University, Ankara, Turkiye
		} \\
		{Email: \{oatopal, cavdar\}@kth.se, ozlemtugfedemir@bilkent.edu.tr 
	} }

    \thanks{This work has been part of 6G-SUSTAIN: Sensing Integrated Elastic 6G Networks for Sustainability project funded by Vinnova in Sweden, and partially funded by the project ``Celtic-Next project RAI-6Green: Robust and AI Native 6G for Green Networks'' with project-id: C2023/1-9 also by Vinnova in Sweden. This work was supported by the Swedish Innovation Agency  (VINNOVA) through the SweWIN center (2023-00572). 
    
    The work of \"O. T. Demir was carried out within the scope of the project 122C149 – Intelligent End-to-End Design of Energy-Efficient and Hardware Impairments-Aware Cell-Free Massive MIMO for Beyond 5G. \"O. T. Demir was supported by the 2232-B International Fellowship for Early Stage Researchers Programme funded by the Scientific and Technological Research Council of Türkiye (TÜBİTAK).}
    
}

\IEEEoverridecommandlockouts
\IEEEpubid{\makebox[\columnwidth]{979-8-3195-4420-9/26/\$31.00~\copyright2026 IEEE \hfill}
\hspace{\columnsep}\makebox[\columnwidth]{ }}

\maketitle

\begin{abstract}
Cell-free massive MIMO promises uniformly high performance by combining densely distributed radio units, coherent transmission, and centralized processing. Unlike earlier radio generations, it depends on dense fronthaul connectivity and a virtualized cloud-RAN architecture. In this setting, energy use is no longer driven primarily by active radio components; instead, fronthaul and processing play a dominant role, calling for a fresh perspective on what defines energy efficiency. This work introduces a modular power model that captures the interplay between radios, fronthaul, and cloud processing. The analysis highlights how design choices, such as functional splits and precoding strategies, shape both fronthaul data load and total power consumption. Centralized precoding provides stronger performance with less resource utilization, while flexible activation of radios and processing elements avoids unnecessary overhead. Overall, the energy efficiency of cell-free massive MIMO grows as antennas are more densely distributed across the coverage area, particularly when combined with end-to-end resource allocation.
\end{abstract}

\section{Introduction}

Cell-free massive MIMO has emerged as a key paradigm for delivering uniformly high performance across the coverage area. This capability stems from the dense deployment of distributed radio units (RUs) and their ability to perform coherent joint transmission/reception \cite{cell-free-book}. To support this operation efficiently, the radio network must be highly centralized: multiple RUs jointly serve the same user equipment (UE), and their baseband processing is executed once in a shared cloud infrastructure. This naturally requires cloud-based processing and network function virtualization, enabling flexible and scalable coordination across the system \cite{ozlem_jsac}. Achieving coherent transmission, however, requires tight synchronization and low-latency interaction between distributed radios and centralized processing. Such requirements can only be met with low-layer functional splits, i.e., options 8, 7.1, and 7.2, which place stringent demands on the fronthaul data rate \cite{3GPP_functional_split}. This represents a fundamental shift from conventional cellular architectures, where baseband processing is largely distributed, and fronthaul requirements are more relaxed. 

This architectural shift brings new challenges, particularly in understanding the dominant sources of power consumption. In contrast to conventional systems, where transmit power is often the primary concern, cell-free massive MIMO introduces significant energy costs in fronthaul transport and cloud processing \cite{ozlem_jsac}. Prior work on energy efficiency in cell-free massive MIMO has primarily focused on optimizing specific parts of the system rather than adopting an end-to-end perspective. For example, \cite{minimize_energy_cf} focus solely on radio resource optimization, while \cite{GroupSparsePrecGreen} considers radio and fronthaul resources. 
However, these works still overlook the role of cloud processing and the full system interaction. \cite{ozlem_jsac} and \cite{energyjournal} propose end-to-end resource orchestration algorithms considering optical fiber and wireless fronthaul, but provide a limited understanding of the energy efficiency trends of cell-free massive MIMO networks.

In this work, we address this gap by utilizing a modular power consumption model that captures transmit power, active radio components, optical fronthaul links, and processing resources. We analyze how functional split choices impact fronthaul load, showing that lower-layer splits significantly increase fronthaul data rate requirements. We further demonstrate how centralized precoding can reduce the number of active antenna ports while implicitly compressing fronthaul traffic, and how dynamic activation of radios and processing resources improves overall efficiency. Finally, we show that the energy-saving potential of cell-free massive MIMO increases with denser antenna deployments when combined with coordinated, end-to-end resource allocation.

\section{Cell-Free Massive MIMO System Model}\label{sec:system}

A downlink cell-free massive MIMO system with time-division duplex and OFDM is considered. $L$ RUs each equipped with $M$ antennas, and $K$ single-antenna UEs are considered. The carrier and sampling frequencies are $f_c$ and $f_s$, respectively.  The total number of subcarriers is $N_{\rm DFT}$ across the total bandwidth of $B$\,Hz, where the subcarrier spacing is denoted by $\Delta f$. $N_{\rm DFT}$ is also the dimension of the discrete Fourier transform (DFT), while the number of used subcarriers is $N_{\rm used}\leq N_{\rm DFT}$. Each OFDM symbol has a duration of  $T_{ s}$ seconds. We assume a block-fading channel model, in which the channels are constant time-invariant and frequency-flat in each coherence block that consists of $N_{\rm smooth}$ consecutive OFDM subcarriers and  $N_{\rm slot}$ OFDM symbols. The channel can be assumed constant across $\tau_c=N_{\rm smooth}N_{\rm slot}$ channel uses, which is the number of useful samples in each coherence block, and takes independent realizations between different blocks \cite[Remark 2.1]{cell-free-book}. 
 
Each coherence block is divided into an uplink training phase with $\tau_p$
samples and a downlink payload data transmission phase with $\tau_d =\tau_c - \tau_p$ samples. %All the UEs are served on the same time-frequency resources using spatial multiplexing. % Moreover, channel estimation and precoding are implemented in each coherence block in the same way. Hence, we will focus on an arbitrary time-frequency resource block as in \cite{cell-free-book}.
We let $\vect{h}_{kl}\in \mathbb{C}^{M}$ denote the frequency-domain channel from UE $k$ to RU $l$ in an arbitrary coherence block. The channels are modeled using uncorrelated Rayleigh fading, i.e.,  $\vect{h}_{kl}\sim\mathcal{N}_{\mathbb{C}}({\bf 0},\beta_{kl}{\bf I}_{M})$ and they are independent for different UEs and RUs. The average channel gain, $\beta_{kl}$, depends on large-scale effects such as geometric attenuation and shadowing. 
To simplify the performance analysis and eliminate operational performance degradation effects, we will consider the following special case. To eliminate interference between the RU-UE channels, we assume that orthogonal waveforms are used for each UE, with transmit power equally divided among them. In such a case, the received signal at an arbitrary UE $k$  can be simplified by  
\begin{equation}
    y_k = \sum_{l=1}^L \vect{h}^{\Ttran}_{kl}\vect{w}_{kl} x_{kl} \sqrt{\rho_{kl}} \varsigma_k + n_k,
\end{equation}
where $\varsigma_k\in \mathbb{C}$ is the complex message symbol of UE $k$, $n_k \sim \mathcal{N}_{\mathbb{C}}({0}, \sigma^2) $ is the additive white Gaussian noise, and $\rho_{kl}$ is the allocated power for UE $k$ by RU $l$, where each RU has limited transmit power, $\sum_{k=1}^{K} \rho_{kl} \leq P_t$. The binary UE-RU association variable, $x_{kl}$, is equal to one if UE $k$ is served by RU $l$. In this special case, maximum ratio transmission (MRT), $\vect{w}_{kl} = \frac{\vect{h}^*_{kl}}{\|\vect{h}_{kl}\|}$, is the optimal unit norm precoding at RU $l$. Also, the power can be distributed among all UEs equally, where $\rho_{kl} = P_t/K$, for all $l,k$. An upper bound on the $k$th UE's ergodic data rate is obtained by Jensen's inequality: \begin{equation}
    R_k \leq B \log_2\left(1+   \gamma \mathbb{E}\left[ \left(\sum_{l=1}^{L} x_{kl}\|\vect{h}_{kl}\| \right)^2 \right] \right), 
\end{equation}
where $\gamma = P_t/(K\sigma^2)$. Using large-array approximation, the effective downlink signal-to-noise ratio (SNR) of UE $k$ can be approximated as\footnote{When the channels are pure LOS, the SNR given \eqref{eq:snr} is the exact one. }
\begin{equation} \vspace{-1mm}
    \mathrm{SNR}_k \approx  \gamma \left(\sum_{l=1}^{L} x_{kl} \sqrt{\beta_{kl}   M_l } \right)^2,\label{eq:snr} \vspace{-1mm}
\end{equation}
where $R_k \approx B \log_2(1+ \mathrm{SNR}_k)$, and $M_l$ is the active number of antennas at RU $l$.

\section{Architecture Overview}\label{sec:architecture}

\begin{figure}[tb]
    \centering
    \includegraphics[width=0.9\linewidth]{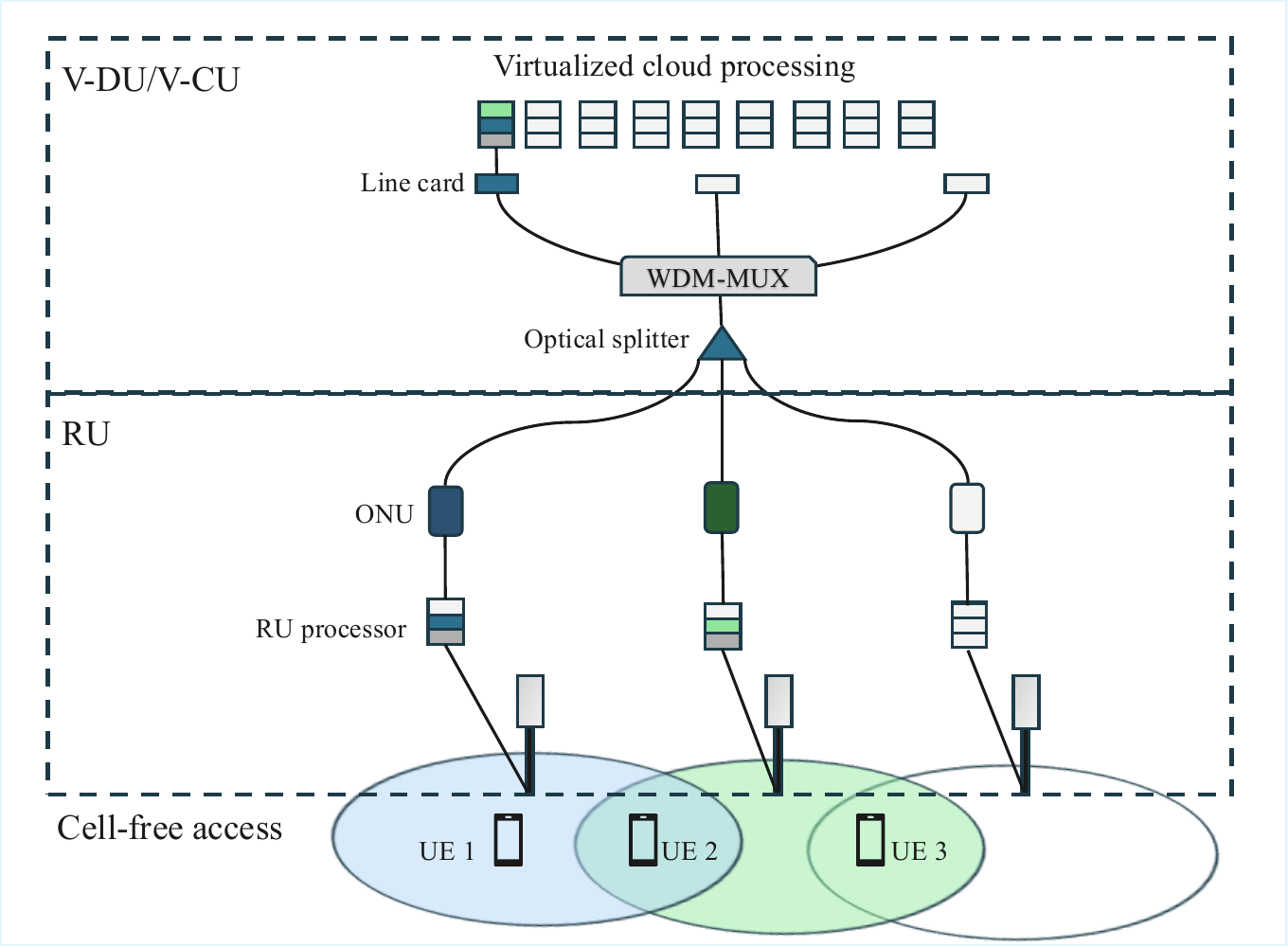}
    \caption{Architectural viewpoint of the cell-free massive MIMO networks.}
    \label{fig:architecture}
\end{figure}

As shown in Fig. \ref{fig:architecture}, a cell-free massive MIMO network is envisioned with a virtualized-RAN deployment. Since we focus on the impact of fronthaul and baseband processing, virtualized central unit (V-CU), backhaul, and core functions, and their effect on power consumption is neglected. In the figure, three UEs are served by two RUs, while the third radio is shut down to save energy. In a virtualized cloud environment, there are $W$ stacks of general-purpose processors (GPPs). These pooled GPPs handle baseband processing thanks to their computational power and flexibility. The workload for each GPP is assigned through a dispatcher managed by a global cloud controller \cite{sigwele2017energy}. The same colors are used to show which RUs are connected to which GPP. Virtualization allows GPP-1 to process the loads from RU-1 and RU-2 jointly, allowing other processors to be turned off.

Evolved CPRI (eCPRI) specification is considered for the fronthaul transmission. A time- and wavelength-division multiplexed passive optical network (TWDM-PON) is employed as the fronthaul transport network to carry eCPRI packets, ensuring the high-capacity fronthaul links in a cell-free massive MIMO network \cite{wang2017virtualized}. Each RU is linked to an optical network unit (ONU) that operates on one of several wavelengths in the fiber network. Multiple RUs can share a single wavelength through time-division multiplexing. Within the virtualized cloud, an optical line terminal (OLT) is equipped with a wavelength-division multiplexing unit (WDM MUX) and several line cards (LCs), each connected to a GPP. Since each LC supports only one wavelength, the signals of a given RU are processed by the GPP operating on the same wavelength.

\subsection{Functional Splits and Fronthaul Load}

Functional split refers to how baseband processing tasks are divided between the RUs and centralized cloud-based processors, determining where different processing functions are executed.
Cell-free massive MIMO relies on coherent joint transmission, which can only be realized by low-PHY functional split options, i.e., options $8$, $7.1$, and $7.2$ \cite{3GPP_functional_split}.  Although higher split options reduce the fronthaul data rate, they constrain phase synchronization among RUs and prevent coherent joint transmission. The fronthaul rate requirement of an RU depends on the chosen functional split. For option $8$, all IQ data need to be transmitted over fronthaul, where the rate requirement becomes 
\begin{equation}
    \bar{R}_{8} = 2 \Delta f N_{\mathrm{DFT}} N_{\mathrm{bits}} M,
\end{equation} where $N_{\mathrm{bits}}$ is the number of quantization bits.

In the split option $7.1$, the filtering and DFT are realized at the RU-site, lowering the fronthaul rate requirement only with active data subcarriers:
\begin{equation}
    \bar{R}_{7.1} = 2 \Delta f N_{\mathrm{used}} N_{\mathrm{bits}} M.
\end{equation}
 Since $N_{\mathrm{used}} \leq N_{\mathrm{DFT}}$, we can guarantee $\bar{R}_{7.1} \leq \bar{R}_{8}$. In both cases, the fronthaul rate scales with the number of active antenna elements at the RU, which will be observed to create a significant performance bottleneck in the numerical analysis.

In the split option $7.2$, filtering, DFT, mapping, and precoding are realized at the RU-site, making the fronthaul rate requirement to scale with the number of data streams:
\begin{equation}
    \bar{R}_{7.2} = 2 \Delta f N_{\mathrm{used}} N_{\mathrm{bits}} |\mathcal{K}|,
\end{equation}
where $|\mathcal{K}|$ is the number of UEs that are served by the RU. In cell-free massive MIMO, RUs can serve UEs with different sets and set sizes, so $|\mathcal{K}|$ can be different for each RU.

\begin{remark}
    Split option $7.2$ enforces precoding to be implemented locally in the RU-site; therefore, distributed precoding schemes must be implemented in the cell-free access. 
\end{remark}

\begin{figure}[tb]
    \centering
    \includegraphics[width=\linewidth]{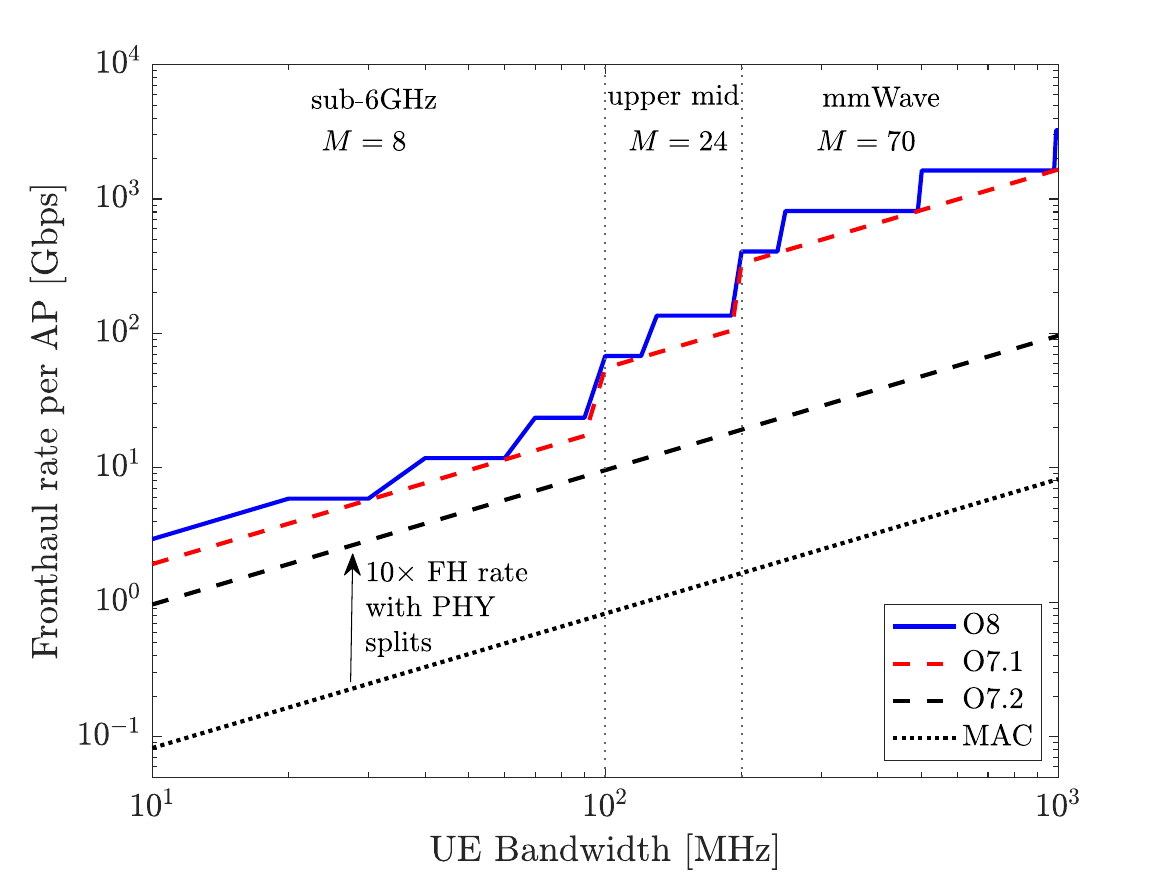}
    \caption{Fronthaul rate requirement under different functional splits per AP vs UE bandwidth. MAC split cannot be realized in cell-free massive MIMO, included just as a reference.}
    \label{fig:SingleAPFronthaulBandwdith}
\end{figure}

\begin{table}[tb]
\centering
\caption{System parameters for Fig. \ref{fig:SingleAPFronthaulBandwdith}.}
\footnotesize
\begin{tabular}{|l|l|l|l|}
\hline
Param. & sub-6GHz & upper-mid & mmWave \\ \hline
$f_c$ & $5\,$GHz & $15\,$GHz & $30\,$GHz \\ 
$B$ & $10-100\,$MHz & $0.1-0.2\,$GHz & $0.2-1\,$GHz \\ 
$\Delta f$ & $30\,$kHz & $60\,$kHz & $240\,$kHz \\ 
$M$ & $8$ & $24$ & $70$ \\  \hline
\end{tabular}%
\label{tab:params}
\end{table}

5G deployment depends on massive MIMO systems, for which the fronthaul design and its limitations cannot be directly applied to cell-free massive MIMO. Cell-free massive MIMO relies on coherent transmission by densely deployed low-cost RUs, which changes the fronthaul rate significantly under the PHY splits, as shown in Fig. \ref{fig:SingleAPFronthaulBandwdith}. In this setup, $N_{\rm bits}=12$ is considered. The remaining parameters change based on the chosen band, and are given in Table \ref{tab:params}. In deciding the number of antennas, the antenna aperture is kept equal to combat the increasing path loss effect with the higher bands.  In MAC layer splits, the fronthaul rate scales with the total rate of the UEs served by the RU, $\sum_{k=1}^{K}x_{kl} R_k$. The current literature on cell-free massive MIMO fronthaul assumes the MAC layer fronthaul rate, although it is not implementable. As can be seen from Fig. \ref{fig:SingleAPFronthaulBandwdith}, this assumption significantly underestimates the fronthaul limitations, as the  PHY splits require more than 10 times the fronthaul rate compared to the MAC splits. As expected, split 8 requires the highest fronthaul rate. It increases as a step function since it scales with $N_{\mathacr{DFT}} = 2^{\lceil\log_2(N_{\mathacr{used}})\rceil}$. The required rate ramps up to $1\,$Tbps as both higher bandwidths and a higher number of antennas are required.  Lowering fronthaul quantization levels, $N_{\rm bits}$, or implementing an array of subarray structures seems necessary to lower these high-rate requirements on the higher bands, reducing the rate performance due to the hardware distortions.

\begin{figure}[tb]
    \centering
    \includegraphics[width=\linewidth]{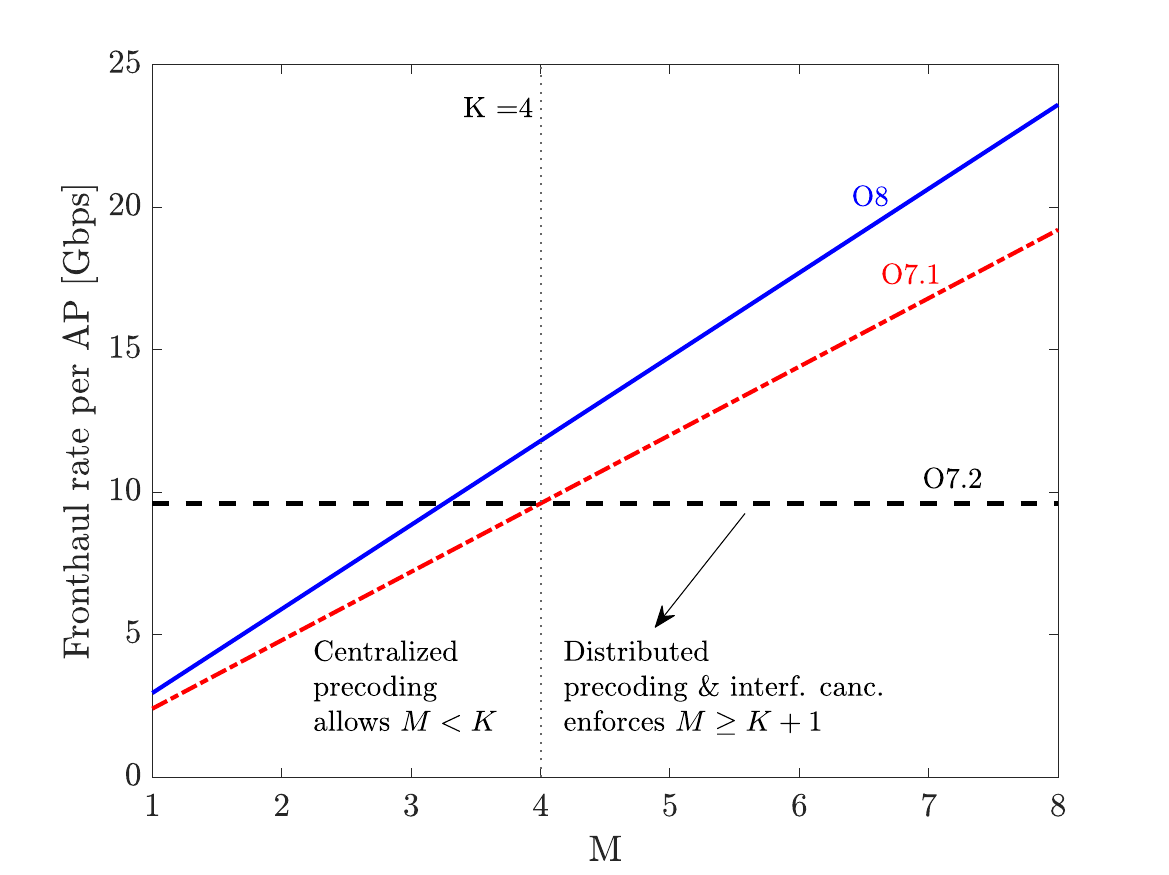}
    \caption{Fronthaul rate requirement under different functional splits per AP vs number of active antennas. UE bandwidth is equal to $100\,$MHz.}
    \label{fig:SingleAPFronthaulAntenna}
\end{figure}

Fig. \ref{fig:SingleAPFronthaulAntenna} illustrates the fronthaul rate requirement for a varying number of antennas at the RU. Cell-free massive MIMO offers a unique advantage: the number of active antennas at an RU can be less than the number of data streams or UEs it serves. This advantage is only harnessed by centralized precoding schemes, which necessitate split options $8$ and $7.1$. In distributed precoding, each RU tries to cancel out the interference of the UEs it serves, inherently requiring either more RUs or more antennas to be activated. Therefore, as shown in Fig. \ref{fig:SingleAPFronthaulAntenna}, the split options $8$ and $7.1$ can lower the fronthaul rate per RU compared to option $7.2$.

\subsection{Virtualization and Processing}
\begin{figure}[tb]
    \centering
    \includegraphics[width=\linewidth]{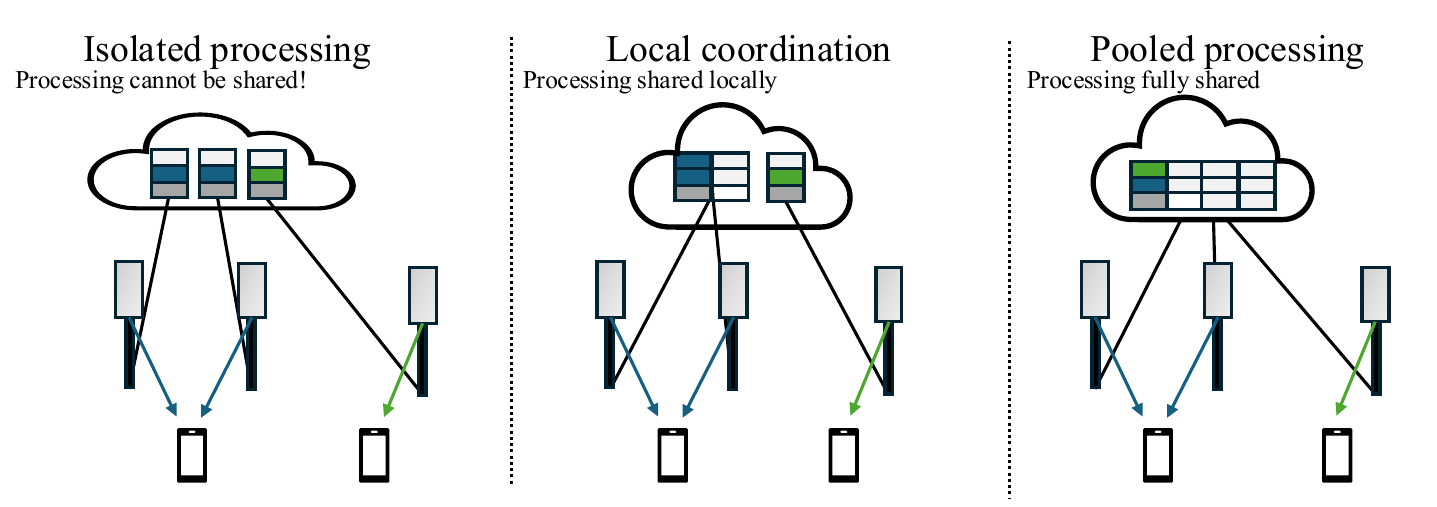}
        \vspace{-0.7cm}
    \caption{The impact of RAN evolution on processing sharing capability for cell-free massive MIMO systems. Virtualization and cloudification allow sharing the processing load of several cells, leading to higher resource efficiency.}
    \label{fig:Processing}
\end{figure}

Based on the chosen functional split option, some low-PHY baseband operations will be carried out at the RU site, and the remaining operations will be carried out at the cloud site. Network function virtualization enables baseband operations to be implemented on GPPs. Each GPP has an idle power consumption that can be high even if the network load is low. Virtualization enables software and hardware upgrades when needed, thereby improving resource efficiency. However, it does not necessarily mean processing resource sharing. Fig. \ref{fig:Processing} demonstrates different levels of processing resource sharing. In all methods, we consider a centralized cloud, which is necessary for cell-free operation, where several RUs usually serve the same set of UEs. However, isolated processors can be assigned to each RU, which means even if the processing load for a single RU is not high, it consumes the idle power of that processor. As the processing is pooled efficiently, they can be shared among RUs, which can allow shutting down more GPPs. 

Gigabit-operations-per-second (GOPS) is the measure of required operations, and the calculation 
 of the GOPS for the operations considered in this work are given in Table \ref{tab:GOPS_table} \cite{energyjournal}. $\mathrm{W}_r$ and $\mathrm{SE}_r$ denote the ratio of the bandwidth and the ratio of the spectral efficiency (SE) of a UE for this work to the reference setup \cite{Debaillie2015a}. In the reference setup, $20$\,MHz bandwidth is chosen, and the SE is equal to $6$\,bit/s/Hz. 

\begin{table}[tb]
\centering
\caption{GOPS formulations for various operations and their execution locations under different functional splits.}
\footnotesize
\vspace{-2mm}
\resizebox{\linewidth}{!}{%
\begin{tabular}{|l|l|l|l|l|l|}
\hline 
Function & GOPS per unit* & Factor & 8 & 7.1 & 7.2 \\ \hline
$C_{\mathrm{filter}, l}$ & ${40 f_s}/{10^9}$ & $M_l$ & Cloud & RU & RU \\ 
$ C_{\mathrm{DFT}, l}$ & $\frac{8 N_{\mathrm{DFT}} \log_2(N_{\mathrm{DFT}})}{T_s10^9}$ & $M_l$ & Cloud & RU & RU \\ 
$C_{\mathrm{map},l}$ & $ 1.3 \mathrm{W}_r \mathrm{SE}^{1.5}_r $ & $\mathacr{X}_l$ $^{\dagger}$ & Cloud & Cloud & RU \\
$ C_{\mathrm{prec}, l}$ & $\left(\frac{8 \tau_d N_{\mathrm{used}}}{T_s 10^9 \tau_c} \right)$ & $ M_l \mathacr{X}_l$ & Cloud & Cloud & RU \\    
$C_{\mathrm{mod},l}$ & $ 1.3 \mathrm{W}_r $ & $M_l$ & Cloud & Cloud & Cloud \\
$C_{\mathrm{cod},l}$ & $ 5.2 \mathrm{W}_r \mathrm{SE}_r $ & $ \mathacr{X}_l$ & Cloud & Cloud & Cloud \\
$C_{\mathrm{netw},l}$ & $ 8 \mathrm{W}_r \mathrm{SE}_r $ & $1$ &Cloud & Cloud & Cloud \\ \hline  
\end{tabular}% 
}
\vspace{0.2mm}

\footnotesize{ *Total GOPS calculated by multiplying GOPS per unit and unit factor. \\ $^{\dagger}$ $\mathacr{X}_l = \sum_{i=1}^K x_{il}$ is defined for brevity. }
\label{tab:GOPS_table}
\end{table}

\section{Radio, Processing, and Fronthaul Effects in Power Consumption}
\label{sec:power_consumption}
In this section, we model network power consumption for the downlink operation. The network power consumption can be calculated as 
\begin{equation}
 P_{\mathrm{tot}} = \sum_{l=1}^{L}  P_{\mathrm{RU},l}   +    P_{\mathrm{Cloud }}, 
\end{equation}
where $P_{\mathrm{RU},l} $ is the power consumed at RU $l$, including radio hardware, transmit power, processing, and fronthaul. $P_{\mathrm{Cloud }}$ is the power consumed at the cloud  \cite{ozlem_jsac}. The power consumption of the backhaul and the core is ignored in this work, since they have a negligible effect compared to the radio and processing \cite{milano}. 

\subsection{Radio-Site Power Consumption}
Power consumption at the RU-site can be categorized under three main factors: $(1)$ transmit and hardware power consumption for the access channel; $(2)$ power consumption for processing done at the RU-site (depends on the chosen functional split), and $(3)$  power consumed for the fronthaul. The power consumption of RU $l$  becomes 
\begin{equation}
\begin{aligned}
    P_{\mathrm{RU},l} = & P^{\mathrm{hard}}_{\mathrm{RU},l} + P^{\mathrm{proc}}_{\mathrm{RU},l} + P^{\mathrm{frth}}_{\mathrm{RU},l}.
    \end{aligned}
\end{equation}
In hardware power consumption, we constitute both hardware-dependent static power consumption, and the load-dependent total transmit power: 
\begin{align}
    P^{\mathrm{hard}}_{\mathrm{RU},l} = M_l P_{\mathrm{st}}   + \Delta^{\rm tr} P_t
\end{align}
where $P_{\mathrm{st}}$ is the static power consumption per active RF chain and $\Delta^{\rm tr}\geq 1$ is the slope of the load-dependent transmit power consumption.
$P_{\mathrm{RU},l}^{\mathrm{proc}}$ is the power consumption by the processing done at RU $l$, and it depends on the chosen functional split, and is calculated as  
\begin{equation}
P^{\mathrm{proc}}_{\mathrm{RU},l} = \frac{1}{\sigma^{\mathrm{RU}}_{\mathrm{c}}} \left((1-\mathacr{I}_{8}) P^{\mathrm{proc}}_{\mathrm{RU},0} +   \Delta_{r}\frac{ C_{\mathrm{RU},l} }{ C_{\mathrm{RU}}^\mathrm{max}}\right),
\end{equation}
where $P^{\mathrm{proc}}_{\mathrm{RU},0}$ is the idle processing power, and $C_{\mathrm{RU},l}$ is the GOPS at the RU-site. The value of $C_{\mathrm{RU},l}$ can be calculated by summing the processes given in Table \ref{tab:GOPS_table} marked by RU for chosen functional split. $\mathacr{I}_{\mathcal{X}}$ is a binary variable that is equal to one for split option $\mathcal{X}$, and zero for other split options. $ C_{\mathrm{RU}}^\mathrm{max}$ is the maximum processing capacityy of the RU in GOPS. $0 < \sigma^{\mathrm{RU}}_{\mathrm{c}} \leq 1$ is the cooling efficiency at any RU. $\Delta_{r}$ is the slope of the load-dependent part.

$P^{\mathrm{frth}}_{\mathrm{RU},l}$ is the fronthaul power consumption, which is equal to the power consumption of the ONU,  $P^{\mathrm{frth}}_{\mathrm{RU},l}= P_{\mathrm{ONU}}$.

\subsection{Cloud-Site Power Consumption}
The total power consumption in the cloud site can be given as 
\begin{align} \label{eq:CPU-power-consumption}
    P_{\rm Cloud} =&P_{\rm fixed}+  P^{\mathrm{frth}}_{\rm Cloud} + P^{\mathrm{proc}}_{\rm Cloud},
  \end{align}  
where $P_{\rm fixed}$ is the load-independent fixed power consumption that includes the power consumption of the cloud dispatcher, housing facilities, etc. 

The processing power consumption can be given as 
    \begin{align}
    P^{\mathrm{proc}}_{\rm Cloud} = 
    \frac{P_{{\rm GPP},0}^{\rm proc} W_a +\Delta_{c} \frac{C_{\rm GPP}}{C_{\rm GPP}^{\rm  max}}}{\sigma_{\rm c}^{\rm Cloud}},
    \end{align}

The cooling efficiency is $0<\sigma_{\rm c}^{\rm Cloud}\leq1$. $P_{{\rm GPP},0}^{\rm proc}$ is the idle power consumption of a GPP.  $\Delta_c$ is the slope of the load-dependent processing power consumption of each GPP. For each GPP, the maximum processing capacity is given by $C_{\rm GPP}^{\rm max}$ in GOPS. The total processing utilization is given by $0\leq C_{{\rm GPP}} \leq W C_{\rm GPP}^{\rm  max}$ in GOPS, which is  $C_{{\rm GPP}} = \sum_{l=1}^{L}z_l C_{l, \rm GPP}$ can be calculated by summing all GOPS that will be implemented in the cloud-site given in Table \ref{tab:GOPS_table} only for RU $l$.  $W_a$ is the number of active GPPs, and it depends on the processing load at the cloud and the processing sharing ability. For isolated processing $$W^{\mathrm{iso}}_a = \sum_{l=1}^L z_l,$$ which is the number of active RUs. For the local coordination, $$W^{\mathrm{loc}}_a = \sum_{n=1}^{N}  \left \lceil  \frac{\sum_{l\in \mathcal{L}_n} z_l C_{l, \rm GPP}}{C_{\rm GPP}^{\rm  max}} \right \rceil,$$
where $\mathcal{L}_n$ is the $n$th group of RUs that are sharing the processing resources. For the pooled processing, all RUs share all GPPs in the cloud, then the active GPPs become: 
$$W^{\mathrm{pool}}_a =  \left \lceil  \frac{C_{\rm GPP}}{C_{\rm GPP}^{\rm  max}} \right \rceil.$$

The fronthaul power consumption in the cloud site can be given as
    \begin{align}
    P^{\mathrm{frth}}_{\rm Cloud} = 
    \frac{P_{\rm OLT}W_a }{\sigma_{\rm c}^{\rm Cloud}},
    \end{align}
where $P_{\rm OLT}$ is the power consumption of an active optical LC. It is worth mentioning that the LC of an active GPP may be inactive if the corresponding GPP participates only in the processing that is redirected to it from other GPPs.

\section{Energy-Aware Joint Resource Allocation}
In this section, we analyze the energy savings with joint radio, cloud, and fronthaul resource allocation.

We consider a square area of size  $1 \times 1~\text{km}^2$ with a grid-type RU deployment.    If not specified, we consider $L=16$ and $M = 8$. We consider $3$\,GHz for the access links. Uncorrelated Rayleigh fading is assumed in the access channel. The shadowing effect in the access channel is modeled as in \cite{ozlem_jsac}.  We consider 5G and beyond access channel properties, as given in Table \ref{tab:simulation_params}. The optical fronthaul and processing values are taken from \cite{ozlem_jsac}. The UEs are distributed uniformly in the considered area. We run $1000$ Monte Carlo simulations and take the average of the performance results. 

\begin{table}[tb!]
\footnotesize
\caption{Simulation parameters}
\begin{tabular}{|l|l|l|l|}
\hline
$f_s$, $B^{\rm ac}$ & $122.88$, $100$\,MHz   &  $T_s$ & $35.68\,\mu$s \\ \hline
$P_t$, pilot pow.  & $5$, $0.5$\,W  & $P_{\mathrm{fixed}}$  & $120$\,W  \\ \hline
$\sigma^{\mathrm{Cloud}}_{\mathrm{c}}$, $\sigma^{\mathrm{RU}}_{\mathrm{c}}$     & $0.9$, $1$  & $\tau_c$, $\tau_p$  &  $260$, $6$  \\ \hline
$C_{\rm GPP}^{\max }$, $C^{\max }_{\mathrm{RU}}$ & $360$, $180$ GOPS &  $P_{\mathrm{st}}$ & $6.8$\,W \\ \hline
 $P_{\mathrm{OLT}}$, $P_{\mathrm{ONU}}$ & $20$, $1.8$\,W & $\Delta_{r}, \Delta_{c}$ & $74$\,W\\ \hline
$N_{\mathrm{DFT}}$,$N_{\mathrm{used}}$ & 4096, 2667 & $N_{\mathrm{bits}}$ & 12  \\ \hline
$P^{\mathrm{proc}}_{\mathrm{RU},0}$, $P_{{\rm GPP},0}^{\rm proc}$  & $20.8$\,W  &$\Delta^{\mathrm{tr}}$ & $4$ \\ \hline
\end{tabular}
\label{tab:simulation_params}
\vspace{-2mm}
%,  %,
\end{table}

In simulation studies, we consider a heuristic resource allocation method to simplify the analysis. Optimal joint resource allocation is an extensive research task, for which optimization-based algorithms and learning-based algorithms are developed in \cite{energyjournal} and in \cite{zilinICC}, respectively. In our setup, we estimate the SNR of UE $k$ with \eqref{eq:snr}. We set a target $\mathrm{SNR} = 5$ level for all UEs. The RUs are ranked based on the channel gains to the UEs from strongest to weakest. One by one, more RUs are associated with the UE until the target SNR level is reached. The RUs that are not chosen for any UE are shut down, i.e., all processing, fronthaul, and radio power consumption, including idle powers, are assumed to be zero. In this way, the sparsest RU-UE association is targeted.

\begin{figure}[tb]
    \centering
    \includegraphics[width=\linewidth]{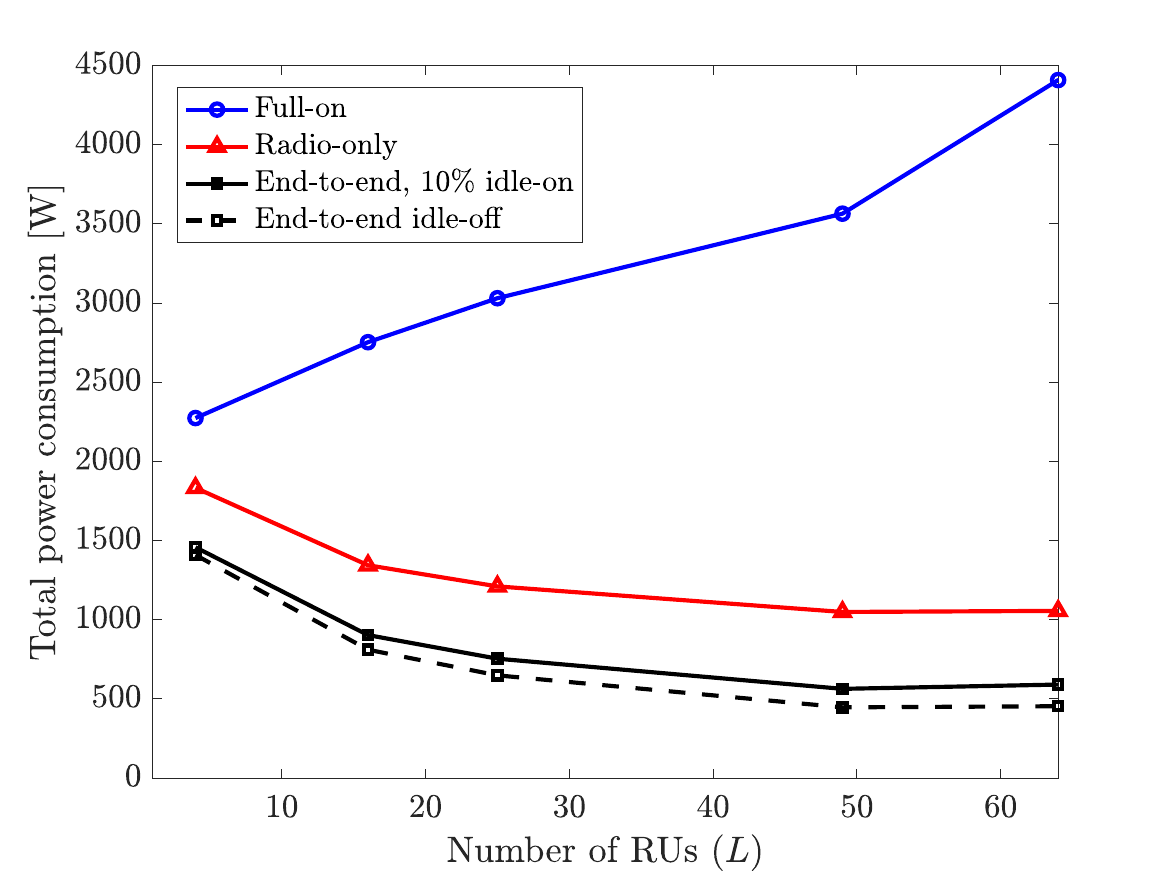}
    \vspace{-0.7cm}
    \caption{The power saving comparison of different resource orchestration mechanisms. As $L$ increases, $M$ decreases in a way to keep $LM$ stable.  Processing pooling, and split option 8 is considered. }
    \label{fig:ResourceOrch}
\end{figure}

Fig. \ref{fig:ResourceOrch} compares the power consumption of different resource allocation methods under varying RU deployments. We set $K=8$ for this figure. As the number of RUs increases, the number of antennas per RU decreases, so that the total number of antennas in the region is kept approximately equal across setups. Full-on is a case where all radio, fronthaul, and cloud processing resources are always on. As the figure shows, denser radio deployment doubles the total power consumption if all network components are active. The radio-only scheme only switches off the radio components, leaving all fronthaul and processing resources at the cloud on. End-to-end resource allocation is the joint radio, fronthaul, and cloud resource allocation, where, based on the activation of the RUs, the unused processors and LCs are also shut down as described in Section \ref{sec:power_consumption}. Both radio-only and end-to-end resource allocation significantly reduce the total power consumption, and also change the power consumption trend. End-to-end resource allocation reduces power consumption by $50\%$ compared to the radio-only power consumption, demonstrating the power contribution of cloud processing and fronthaul in cell-free massive MIMO. Turning off unused network components usually keeps some portion of the power consumption active \cite{3GPP_sleep}. If shutting down equipment reduces the idle power consumption by $90\%$, the energy saving trend seems unchanged, but the power consumption increases by $2\%$.

\begin{figure}[tb]
    \centering
    \includegraphics[width=\linewidth]{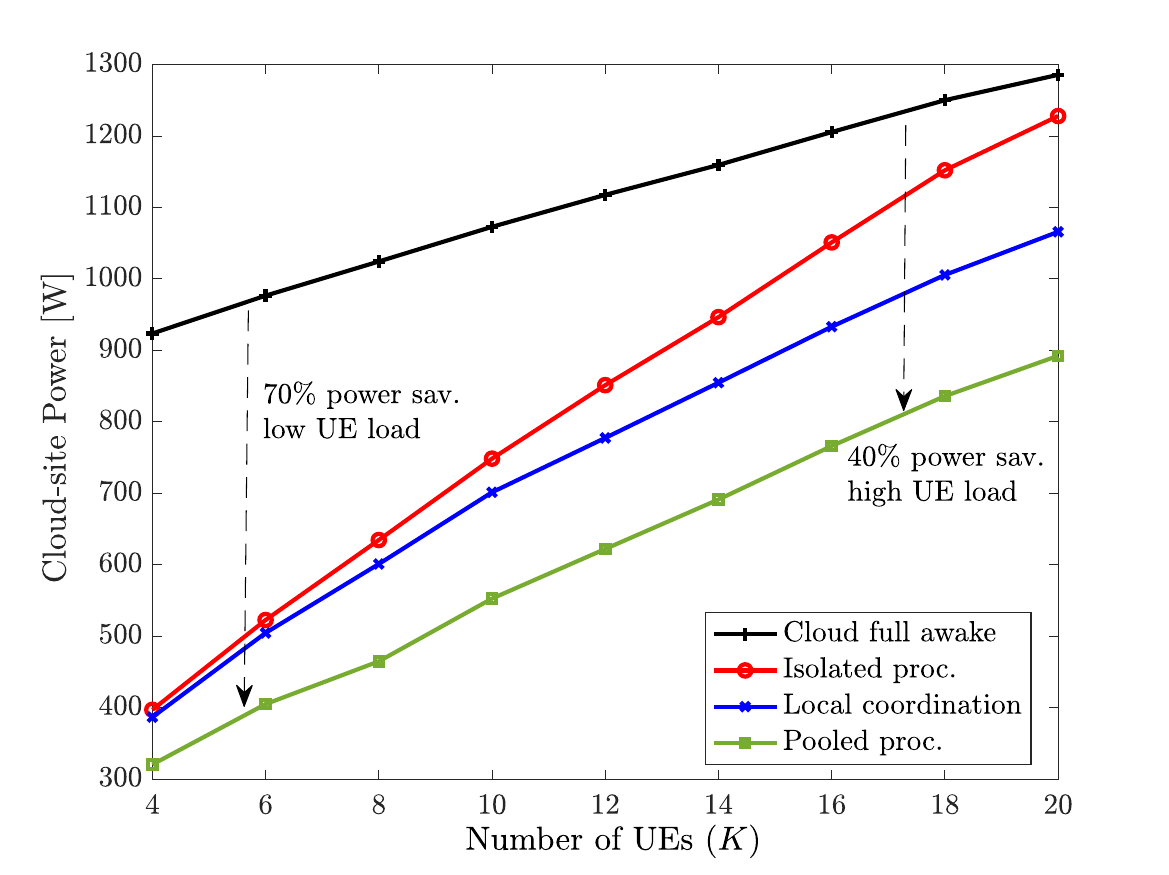}
    \vspace{-0.7cm}
    \caption{The effect of cloud orchestration on power consumption.}
    \label{fig:CloudOrchest}
            \vspace{-0.4cm}
\end{figure}

Fig. \ref{fig:CloudOrchest} shows the cloud-site power consumption with different processing sharing methods. Processing pooling reduces power consumption by $70\%$ and by $40\%$ compared to the full-awake method under the low-load and high-load  scenarios, respectively. As the number of UEs increases, power savings reduce due to increased processing load. 

\begin{figure}[tb]
    \centering
    \includegraphics[width=\linewidth]{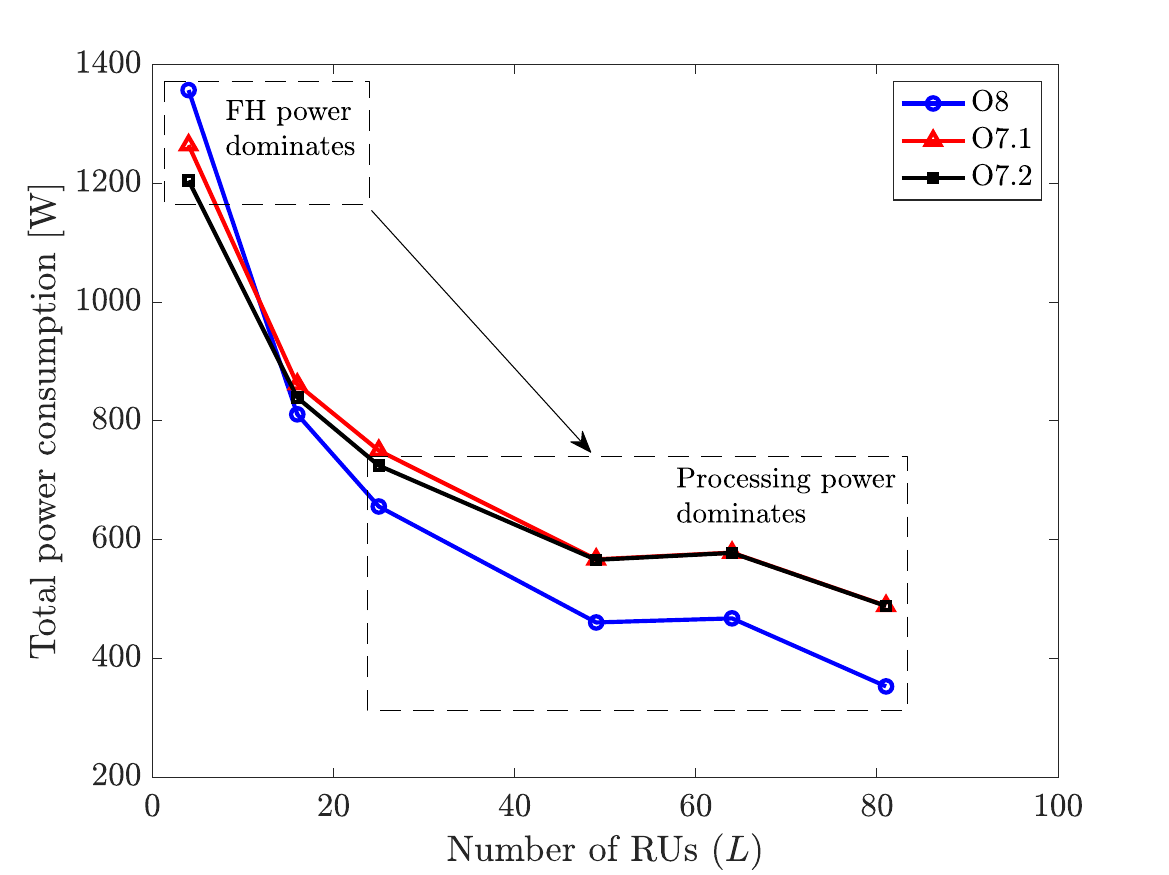}
        \vspace{-0.7cm}
    \caption{The effect of functional splits on power consumption.}
        \vspace{-0.4cm}
    \label{fig:RadioSplits}
\end{figure}

Fig. \ref{fig:RadioSplits} illustrates the impact of the selected functional split on the overall power consumption. Interestingly, when the number of RUs is very small, split option $8$ results in the highest power consumption compared to higher-layer splits. In this regime, whether processing is performed locally or centrally is of limited importance, since an active RU already utilizes a substantial portion of its processing resources. However, the split $8$ significantly increases the fronthaul load, requiring a larger number of active LCs and thereby raising the total power consumption. In contrast, for more distributed radio deployments, centralizing the processing becomes advantageous, as it enables the deactivation of local processors at the RU sites. Moreover, fewer antennas in total are activated in this regime, reducing the total fronthaul load, and the power consumption differences among the various split options become less pronounced.

\section{Conclusion}
\vspace{-1mm}
In this work, we analyzed the impact of different network components on the power consumption of a cell-free massive MIMO network. Due to the dense radio deployment, fronthaul and processing become significant power consumption resources, requiring end-to-end resource allocation. Pooling the processing resources reduces power consumption between $40\%$ and $70\%$ under varying network loads. As the radio deployment becomes more distributed, the processing power consumption dominates, making split $8$ the most power efficient option.

\bibliographystyle{IEEEtran}
\bibliography{IEEEabrv,refs}

\end{document}